# Thermal Expansion and Compressibility in Superconducting $Na_xCoO_2 \cdot 4xD_2O$ (x≈1/3): Evidence for Pressure-Induced Charge Redistribution

J. D. Jorgensen, M. Avdeev, D. G. Hinks, P. W. Barnes, and S. Short
Materials Science Division, Argonne National Laboratory, Argonne, IL 60439

ABSTRACT

We have performed thermal expansion and compressibility measurements on the recently discovered superconducting material $Na_xCoO_2 \cdot 4xD_2O$ (x≈1/3) using neutron powder diffraction over the temperature range 10-295 K and the pressure range 0-0.6 GPa. Pressure measurements were done in a helium-gas pressure cell. Both the thermal expansion and compressibility are very anisotropic, with the largest effects along the c axis, as would be expected for a layered material with weak hydrogen bonding nominally along the c axis. Near room temperature, the anisotropies of the thermal expansion and compressibility of the hexagonal crystal structure are nearly the same [$(\Delta c/c)/(\Delta a/a) \approx 3$-4], with a 100° C change in temperature being roughly equivalent to 0.2 GPa pressure. This would imply that changes in atom position parameters are also the same, but this is not the case. While the effects of temperature on the atom positions are essentially what one might expect, the effects of pressure are surprising. With increasing pressure, the thickness of the $CoO_2$ layer increases, due to the combined effects of an increasing Co-O bond length and changes in the O-Co-O angles of the $CoO_6$ octahedra. We conclude that this unusual effect results from pressure-induced strengthening of the hydrogen bonding between the $Na_x(D_2O)_{4x}$ layers and the $CoO_2$ layers. The strengthening of these hydrogen bonds requires that charge be moved from the somewhere else in the structure; hence, there is a pressure induced charge redistribution that weakens (lengthens) the Co-O bonds and changes the electronic structure of the superconducting $CoO_2$ layers.

PACS: 74.62.Fj, 61.12.Ld, 65.40.De, 62.50.+p

INTRODUCTION

Despite it's low $T_c$ of 4.5 K, the recently discovered superconducting material $Na_xCoO_2 \cdot 4xH_2O$ ($x \approx 1/3$) [1] has generated substantial interest because it appears to be the first superconducting layered metal oxide involving 3d transition metals other than copper. Subsequent measurements and theoretical discussion have raised additional interesting questions, such as the possible role of magnetic ordering on the frustrated triangular cobalt lattice, charge ordering, and possible similarities to the layered copper oxides, including a dome-like behavior of $T_c$ versus chemical composition.[2]

At the same time that these studies of the physics were proceeding, many laboratories were still struggling to understand the synthesis chemistry,[3-7] exact composition,[5, 8-10] crystal structure,[11-15] and other basic properties of the material. This work was made more difficult by the fact that the material is unstable in dry air or at temperatures only a few degrees above room temperature.[3, 10] Thus, some of these basic questions are still a subject of debate. It is in this context that we undertook studies to determine the basic physical properties of thermal expansion and compressibility of superconducting $Na_xCoO_2 \cdot 4xD_2O$ ($x \approx 1/3$). As expected for a layered compound, we observed a large anisotropy in both the thermal expansion and the compressibility. However, refinement of the atom positions for the compressibility data revealed an unexpected and intriguing result. We observed that the thickness of the $CoO_2$ layer actually increases with pressure, as a result of Co-O bond lengthening combined with a decrease in the degree of distortion of the $CoO_6$ octahedra. We conclude that this surprising behavior of the $CoO_2$ layer is a response to a pressure-induced strengthening of the hydrogen bonding in the $Na_x(D_2O)_{4x}$ layer. The strengthening of the hydrogen bonding between the $D_2O$ molecules and the oxygen atoms in the $CoO_2$ layers requires that charge be redistributed from the Co-O bonds to the hydrogen bonds. The broad implication of this result is that superconducting $Na_xCoO_2 \cdot 4xD_2O$ ($x \approx 1/3$) can be viewed as a stacking of superconducting $CoO_2$ layers separated by $Na_x(D_2O)_{4x}$ charge reservoir layers with the potential for charge transfer between the charge reservoir and the supercconducting layers, much like the layered copper oxides.[16, 17]

SYNTHESIS

The measurements reported here were done with the same sample used for the determination of the crystal structure reported in **Ref. 12**. The synthesis is described in the previous paper. A more detailed description of the synthesis has been reported in a recent paper.[10] A powder sample of the anhydrous compound with composition $Na_{0.61(1)}CoO_2$ was made by repeated heating and grinding of a mixture of $Na_2CO_3$ and $Co_3O_4$ in air at 850° C. An initial powder stoichiometry of $Na_{0.7}CoO_2$ resulted in a sample of composition $Na_{0.61(1)}CoO_2$ due to a slight loss of Na during the heating cycles. For the neutron diffraction experiments, a fully deuterated sample was desired in order to avoid the large neutron incoherent scattering from hydrogen, which adds background to the diffraction data. In order to eliminate any possible hydrogen incorporation during the oxidative deintercalation, either from the acetonitrile or residual $H_2O$ in the acetonitrile (as used in other laboratories)[1, 3, 15] we used $Br_2$ in $D_2O$ as the oxidizing medium. About 10 wt% of $Na_{0.61}CoO_2$ powder in $D_2O$ with a 100 to 200% excess of $Br_2$ was shaken overnight in an air-tight pyrex tube. The resulting powder was then separated by filtration. After washing with $D_2O$ to remove excess $Br_2$ and NaBr, the resulting material was kept in a container with a $D_2O$ relative humidity of 100% at room temperature.

This "as-oxidized" deuterated material contained free $D_2O$, as evidenced by the appearance of peaks from ice in a neutron powder diffraction pattern taken at low temperature. In order to remove only the free $D_2O$ and none of the lattice $D_2O$, knowledge of the vapor pressure of $D_2O$ in equilibrium with the higher deuterate phase is required. Approximately 2g of the material was placed in a pyrex vacuum line and pumped to remove air. The partial pressure of $D_2O$ from the hydrate was measured using a capacitance manometer as the $D_2O$ was volumetrically removed by cooling a calibrated glass bulb attached to the vacuum system to liquid nitrogen temperature. To allow the sample to come to equilibrium at each partial pressure of $D_2O$, this cryopumping was done in small steps -- typically a few minutes of pumping followed by equilibration times of up to several hours. All measurements were done at $23.0 \pm 0.2°$ C. In these measurements, a plateau at about 1 kPa indicates the transition to a two-phase deuterate mixture; i.e., a mixture of the higher deuterate phase and the lower deuterate phase reported to have the composition $Na_{0.3}CoO_2 \cdot 0.6D_2O$.[12] A single phase of the higher deuterate with a minimum amount of free $D_2O$ can be made by equilibrating at a $D_2O$ pressure just above 1 kPa. The sample used for the diffraction measurements was equilibrated at 1.2 kPa $D_2O$ pressure overnight. Neutron diffraction measurements at low temperature confirmed that this sample had no observable free $D_2O$.

NEUTRON DIFFRACTION MEASUREMENTS

Using the same sample, neutron powder diffraction measurements were made on the Special Environment Powder diffractometer at Argonne National Laboratory [18] first as a function of temperature and then as a function of pressure at room temperature over a period of about one week. The sample was confirmed to be superconducting at 4.5 K after being unloaded from the high pressure cell.

Two recent studies [10, 19] show that samples of superconducting $Na_xCoO_2 \cdot 4xD_2O$ ($x \approx 1/3$) can show dramatic time dependence of the transition temperature. The sample used for this study, however, showed no time-dependence of Tc. Tc remained at 4.5 K throughout the measurements described here and for several months thereafter. We are currently exploring what variables in the defect concentrations lead to stable samples.

The temperature-dependent measurements were made by sealing the sample in a vanadium can with helium exchange gas and cooling using a closed-cycle helium (Displex) refrigerator. The free volume in the vanadium can was minimized to ensure that the sample would come to equilibrium with its own vapor pressure with no significant loss of water. The can was not allowed to heat above room temperature at any time during the sample loading or mounting on the diffractometer. Data collection required a few hours at each temperature.

Measurements versus pressure at room temperature were then made in a helium gas pressure cell.[20] The sample was loaded as quickly as possible (~ 2 min) to avoid loss of water during handling. Measurements were made at four pressures, including ambient pressure, up to a maximum pressure of 0.6 GPa (the pressure limit for this pressure cell). Data collection required a few hours at each pressure. A second high pressure experiment, at ambient pressure and the maximum pressure of 0.6 GPa, was then done using argon gas as the pressure transmitting fluid. This measurement was done to explore whether helium might enter the crystal lattice at high pressure, changing the composition and crystal structure. The potential for helium intercalation has been demonstrated by high pressure neutron diffraction experiments on $C_{60}$, where small helium atoms are found to incorporate into the crystal lattice at high pressure while larger atoms like argon do not.[21] In the present experiments, helium and argon pressure fluids gave the same result for the compressibility, leading us to conclude that helium intercalation into the $Na_xCoO_2 \cdot 4xD_2O$ ($x \approx 1/3$) lattice does not occur.

Structure refinements were done using the EXPGUI/GSAS Rietveld software [22] with the structural model presented in our previous paper.[12] In this model, the average structure (ignoring the short-range supercell ordering) is refined with the geometry of the $D_2O$ molecule constrained as a rigid body with fixed O-D distances and D-O-D angle while the $D_2O$ molecule position and orientation are refined. A large amount of work on hydrated compounds and ices, including high-pressure structures, shows that constraining the $D_2O$ molecular geometry should be valid under all conditions used in this study.[23, 24] The most challenging issue for structural refinement of data from $Na_xCoO_2 \cdot 4xD_2O$ ($x \approx 1/3$) is that the supercell ordering, with different coherence lengths in the basal plane and along the c axis, leads to a substantial amount of diffuse scattering [12] that can lead to systematic errors in the refined parameters of the average structure. In the following discussion, we point out some of these systematic errors. However, we believe that these systematic errors should not depend on temperature or pressure. Thus, the changes of the refined atom positions as a function of temperature and pressure should be meaningful while the absolute values could be subject to systematic errors significantly larger that the standard deviations from the refinements.

RESULTS AND DISCUSSION

The refined unit cell volumes and lattice parameters as a function of temperature and pressure are presented in **Fig. 1**. These values are expressed as relative changes with respect to ambient conditions (295 K, 1 atm) so that the temperature- and pressure-induced changes can be readily compared. Fortuitously, the changes from 10-295 K are roughly the same as for the 0-0.6 GPa pressure range we can achieve. The plotting scales are chosen to match the changes in cell volume ($\Delta V/V_0$) at ambient conditions; these same plotting scales are then used to compare $\Delta a/a_0$ and $\Delta c/c_0$ as well as other structural parameters presented in subsequent figures.

In Fig. 1, the temperature dependence is fit with an Einstein equation with a single phonon frequency (i.e., single Einsten temperature),

$$\ln\left(\frac{a_i}{a_0}\right) = \frac{c_i \theta_i}{e^{(\theta_i/T)} - 1} \quad (1)$$

where $a_i$ is V, a, or c and the coefficients $a_0$ (the fitted values at ambient conditions; 295 K, 1 atm.), $c_i$, and $\theta_i$ are given in **Table 1**. This simple equation provides an adequate fit to the data, within the error bars, and gives a simple way to estimate the unit cell volume, lattice parameters and thermal expansions at various temperatures. Volume and linear thermal expansions at four temperatures, 295, 200, 100 and 50 K, are given in **Table 2**.

The pressure dependence is fit with a second order polynomial (for example, $V/V_0=1+b_1P+b_2P^2$, where $b_1$ and $b_2$ are the refined coefficients), but the departure from linear behavior is very small over the pressure range studied here. The volume and linear compressibilities at 295 K and 1 atm are given in **Table 3**. The bulk modulus $[=\Delta P/(\Delta V/V_0)]$ at 1 atm is 43.3(7) GPa. This is in excellent agreement with the work of Park et al. **[24]** who reported a bulk modulus of 43(2) GPa based on in situ synchrotron powder diffraction using a diamond anvil cell with Fluorinert as the pressure transmitting fluid.

As expected for a layered compound with weak bonding along the c axis, both the thermal expansion and the compressibility exhibit substantial anisotropy. At 295 K, the anisotropy of the thermal expansion, $[(\Delta c/c_0)/\Delta T]/[\Delta(a/a_0)/\Delta T]$, is 2.9. The anisotropy increases to 6.4 at 100 K. This agrees qualitatively with the results of Takada et al. **[13]**; however, they did not report their thermal expansion data with enough precision to make a meaningful quantitative comparison.

The compression anisotropy, $[(\Delta c/c_0)/\Delta P]/[\Delta(a/a_0)/\Delta P]$, at 1 atm is 4.1(4). This agrees, within the error bars, with a value of 4.5(1.0) reported by Park et al.**[25]** The agreement is perhaps surprising in light of past experience that the measured compression anisotropy of layered compounds can change significantly for different pressure transmitting fluids.**[26]**

The similarity in the thermal expansion and compression anisotropies suggests that the variation of atom positions with decreasing temperature or increasing pressure would also be similar, but this is not the case. In fact, a surprising behavior is seen with increasing pressure. As shown in **Fig. 2**, the $CoO_2$ layer thickness is essentially constant, within the error bars, versus temperature, but increases significantly with increasing pressure. This pressure-induced increase in the layer thickness can be attributed to two changes in the internal structure. First. the Co-O bond length increases with pressure, as shown in **Fig. 3**. Second, the $CoO_6$ octahedra become less distorted, as measured by the O-Co-O angle, as shown in **Fig. 4**. The $CoO_2$ layer thickness, Co-O bond length and O-Co-O angle are defined as shown in **Fig. 5**.

It is important to comment on the systematic errors when temperature- and pressure-dependent data are presented in **Figs. 2, 3, and 4**. The values at 295 K and 1 atm should be the same for the two measurements; yet, they differ by a few standard deviations. The temperature-dependent data are collected in back scattering, while the pressure-dependent data are collected at a scattering angle of 90°. Thus, the temperature-and pressure-dependent data differ in resolution ($\Delta d/d \approx 0.003$ in back scattering and 0.005 at 90°).**[18]** The lower-resolution data taken at pressure are more susceptible to systematic errors from the diffuse scattering resulting from short-range supercell ordering.**[12]** In an attempt to explore the size of this effect, we performed additional refinements in which the major regions of diffuse scattering were excluded. The slopes (vs. pressure) shown in **Figs. 2, 3, and 4** changed as much as one or two standard deviations, but, in all cases, the $CoO_2$ layer thickness and the Co-O bond length showed statistically significant increases with increasing pressure. We conclude that these trends are real and, moreover, that the systematic errors introduced by the diffuse scattering are not pressure dependent.

Although the pressure-induced behavior may, at first, seem surprising, it is consistent with a large block of data on the compression behavior of hydrated compounds and has a straightforward explanation. Structural studies of a large number of hydrated compounds **[23]**, as well as simple hydrogen-bonded systems such as ice **[24]** consistently show that the strong O-H (or O-D) bonds show negligible change with pressure while the much weaker O•••H hydrogen bonds shorten significantly. This shortening (strengthening) of hydrogen bonds with increasing pressure can be so pronounced that many compounds that do not form hydrates at ambient pressure do so at high pressure.**[23]** Clearly, the pressure-induced strengthening of hydrogen bonding requires that a small amount of charge be transferred into the hydrogen bond from somewhere else in the structure. In the case of $Na_xCoO_2 \cdot 4xD_2O$ ($x \approx 1/3$) the observed pressure-induced lengthening of the Co-O bonds is clear evidence that these bonds are weakened as charge is redistributed to strengthen the hydrogen bonds.

Very similar behavior has been observed in the layered materials $Mg(OH)_2$ (brucite),**[27]** $Mn(OD)_2$, and $B-Co(OD)_2$ **[28]** at high pressure. These three isostructural compounds have layers of $MO_6$ octahedra (M=Mg, Mn, or Co) separated along the c axis by hydroxide layers in which hydrogen bonding competes with H-H (D-D) repulsion. Even though a comparatively large compression is observed along the c axis, resulting from dramatic shortening of the hydrogen bonds, the $MO_2$ layers increase in thickness with increasing pressure as a result of a rather large pressure-induced change in the O-M-O angle with increasing pressure leading to less distortion of the octahedra. This pressure-induced decrease of the octahedral distortion is comparable to what we observe in $Na_xCoO_2 \cdot 4xD_2O$ ($x \approx 1/3$). However, in contrast to our results for $Na_xCoO_2 \cdot 4xD_2O$ ($x \approx 1/3$), the M-O bonds in these compounds with the brucite structure shorten slightly with increasing pressure. The pressure-induced changes in octahedral distortion are large enough that the $MO_2$ layers become thicker in spite of the shortened M-O bonds.

It is tempting to use bond valence sum techniques **[29]** to estimate the change in the effective Ço oxidation state, based on the change in the Co-O bond length. Such calculations suggest a change in the Co oxidation state as large as 0.1 electrons over the 0.6 GPa pressure range of this study. This estimate seems suspiciously high in light of the fact that the entire region of superconducting behavior vs. formal Co oxidation state is roughly of the same magnitude.**[10, 30]** We speculate that the validity of the bond valence sum is compromised in the case of high-pressure data because the coefficients assumed to be constants, especially $r_0$, are pressure dependent.**[20]** These coefficients, normally assumed to be constants, must be pressure dependent to achieve overall charge conservation. In the present case, we do not have an estimate for the pressure dependence of $r_0$ for Co-O bonds; thus, the bond valence sum method cannot be applied to estimate the amount of charge transfer. Nevertheless, it is clear that the electronic state of Co must change with pressure in response to the changes if Co-O bond length and O-Co-O angle.

Measurements of the dependence of $T_c$ on pressure [31, 32] and on Co oxidation state [10] can be used to make a rough estimate of the amount of transferred charge. Lorenz et al. [31] reported an unusual non-linear dependence of $T_c$ on pressure, while a more recent measurement by Sushko et al. [32] shows a linear dependence in the pressure range relevant to our measurements with $dT_c/dP$=-0.37 K-GPa$^{-1}$ (essentially the same value reported by Lorenz et al. at ambient pressure). This pressure coefficient of $T_c$ is comparable to that of electron-doped copper-oxide superconductors with similar $T_c$ values. $T_c$ decreases by about 0.2 K over the pressure range of our structural measurements. Given that the entire superconducting dome as a function of Co oxidation state is about 0.1 electrons wide, [10] the amount of charge redistributed in our experiments to 0.6 GPa is on the order of 0.01 electrons per formula unit. This is consistent with the fact that the amount of charge involved in weak hydrogen bonding is very small.

It is important to mention that Rivadulla et al. [33] have reported a similar compression anisotropy in the anhydrous parent compound $Na_{1-x}CoO_2$ and have explained the large c-axis compressibility in terms of a pressure-induced transition from polaronic to itinerant electron behavior. This scenario implies a shortening of Co-O bonds with increasing pressure, contrary to what we observe. Moreover, the hydrated superconducting compound, $Na_xCoO_2 \cdot 4D_2O$ (x≈1/3), is metallic at all pressures. Thus, we conclude that this explanation is not applicable to the present data.

CONCLUSIONS

The results reported here shed new light on the interpretation of measurements of $T_c$ vs. pressure for $Na_xCoO_2 \cdot 4xD_2O$ (x≈1/3). Both Lorenz et al. [31] and Sushko et al. [32] interpret their observations of a negative $dT_c/dP$ in terms of the dependence of $T_c$ on interlayer separation assuming no change in the Co oxidation state. They speculate that superconductivity in this material depends on decreasing the interlayer coupling to give two-dimensional behavior. Our results show that interlayer coupling is not the only variable to be considered when interpreting the effect of pressure on $T_c$. In fact, pressure-induced charge redistribution may be the dominant effect leading to a decrease of $T_c$. Moreover, because increasing pressure results in charge redistribution, removing charge from the Co-O bonds, it is impossible to independently measure the effect of interlayer coupling. Understanding the effects of pressure on $T_c$ in $Na_xCoO_2 \cdot 4xD_2O$ will clearly require additional studies, including measurements on samples with compositions that have reduced $T_c$s, as was done in the layered copper oxides.[34] In those materials, for example it was found that $dT_c/dP$ was often very different for optimally doped (maximum $T_c$), underdoped and overdoped materials. If pressure-induced charge transfer dominates $dT_c/dP$ for $Na_xCoO_2 \cdot 4xD_2O$ (x≈1/3), similar effects should be seen.

The results reported here, have important broader implications for understanding superconductivity in layered $Na_xCoO_2 \cdot 4xD_2O$. We conclude that, like the layered copper oxides,[16,17] this compound should be viewed as an alternating stack of superconducting and charge reservoir layers, with the charge state in the superconducting layers being controlled by the composition and structure of the charge reservoir. The important new information from this study is that the Na content is not the only variable parameter in the charge reservoir layer. The strength of the hydrogen bonding can also lead to significant charge redistribution. This raises the possibility that the water content ( if it can be varied away from the ideal hydrate composition of four water molecules per sodium ion **[12, 35]**), as well as the degree of hydrogen-bond ordering, could be additional factors controlling $T_c$. Such behavior is clearly reminiscent of the layered copper oxides, where it was found that oxygen vacancy ordering, as well as composition, in the charge reservoir layer, could have a significant effect on the superconducting behavior.**[36]**

ACKNOWLEGEMENTS

This work was supported by the U. S. Department of Energy, Office of Science, Division of Basic Energy Sciences, under contract No. W-31-109-ENG-38.

Table 1: Coefficients used in equation (1) and Fig. 1 (see text) obtained by least squares fitting an Einstein equation to the unit cell volume, a axis, and c axis vs. temperature. Numbers in parenthesis are standard deviations of the last significant digit. The fit parameters for $a_0$ are slightly different than the measured parameters given in Fig. 1.

|   | $a_0$ | $c_i$ | $\theta_i$ |
|---|---|---|---|
| V | 135.02(1) Å$^3$ | 5.1(2) x 10$^{-5}$ | 260(20) |
| a | 2.81706(6) Å | 1.2(1) x 10$^{-5}$ | 410(40) |
| c | 19.6455(6) Å | 3.04(5) x 10$^{-5}$ | 206(9) |

Table 2: Volume and linear thermal expansions at 295, 200, 100, and 50 K, as estimated by the fit of equation (1) to the neutron powder diffraction data. Numbers in parenthesis are the standard deviations based on the deviations of the coefficients in Table 1.

|   | 295 K | 200 K | 100 K | 50 K |
|---|---|---|---|---|
| $\Delta(V/V_0)/\Delta T$ | 4.9(2)x10$^{-5}$ K$^{-1}$ | 4.5(2)x10$^{-5}$ K$^{-1}$ | 3.0(3)x10$^{-5}$ K$^{-1}$ | 7(2)x10$^{-6}$ K$^{-1}$ |
| $\Delta(a/a_0)/\Delta T$ | 1.0(1)x10$^{-5}$ K$^{-1}$ | 9(1)x10$^{-6}$ K$^{-1}$ | 3.4(8)x10$^{-6}$ K$^{-1}$ | 2(1)x10$^{-7}$ K$^{-1}$ |
| $\Delta(c/c_0)/\Delta T$ | 2.94(5)x10$^{-5}$ K$^{-1}$ | 2.79(5)x10$^{-5}$ K$^{-1}$ | 2.16(7)x10$^{-5}$ K$^{-1}$ | 8.6(8)x10$^{-6}$ K$^{-1}$ |

Table 3: Volume and linear compressibilities at 295 K and 1 atm obtained from least squares fitting of a second order polynomial to the data shown in Fig. 1. Numbers in parenthesis are standard deviations of last significant digit.

| $\Delta(V/V_0)/\Delta P$ | -0.0231(4) GPa$^{-1}$ |
|---|---|
| $\Delta(a/a_0)/\Delta P$ | -0.0038(3) GPa$^{-1}$ |
| $\Delta(c/c_0)/\Delta P$ | -0.0155(2) GPa$^{-1}$ |

FIGURE CAPTIONS

Fig. 1. Normalized unit cell volume, V, and lattice paratmeters a and c, vs. temperature and pressure as determined from neutron powder diffraction data using Rietveld refinement. The plots are normalized using $V_0$=136.303(6), $a_0$=2.82166(5), $c_0$=19.7681(6) which are the refined values from the temperature dependent data at 295 K. The temperature dependent data are least squares fit with an Einstein equation [see equation (1) and Table 1] and the pressure dependent data are fit with a second order polynomial. For $V/V_0$ the data for the lowest two pressures correspond perfectly with the temperature dependent data for these plotting scales; thus, the round symbols for these data are obscured by the square symbols. Error bars are smaller than the symbols.

Fig. 2. $CoO_2$ layer thickness vs. temperature at 1 atm (dashed line) and pressure at 295 K (solid line) in $Na_xCoO_2 \cdot 4xD_2O$ (x≈1/3).

Fig. 3. Co-O bond length vs. temperature 1 atm (dashed line) and pressure at 295 K (solid line) in $Na_xCoO_2 \cdot 4xD_2O$ (x≈1/3).

Fig. 4. O-Co-O bond angle vs. temperature 1 atm (dashed line) and pressure at 295 K (solid line) in the $CoO_6$ octahedra of $Na_xCoO_2 \cdot 4xD_2O$ (x≈1/3). This angle is a measure of the degree of distortion of the $CoO_6$ octahedra, with an angle of 90° for undistorted octahedra.

Fig. 5. Partial structure of $Na_xCoO_2 \cdot 4xD_2O$ (x≈1/3) showing how Na atoms and $D_2O$ molecules are positioned with respect to the $CoO_6$ octahedra that make up the $CoO_2$ layers. The full unit cell is shown in **Ref. 12**. The $CoO_2$ layer thickness, Co-O bond length, and O-Co-O angle (solid curved line with arrows) referred to in the text and in **Figs. 2, 3, and 4** are defined as shown here. The open dashed line shows the D•••O hydrogen bond that shortens dramatically under pressure, while the solid dashed lines shows the Co-O bonds that lengths with increasing pressure.

Fig. 1(a)

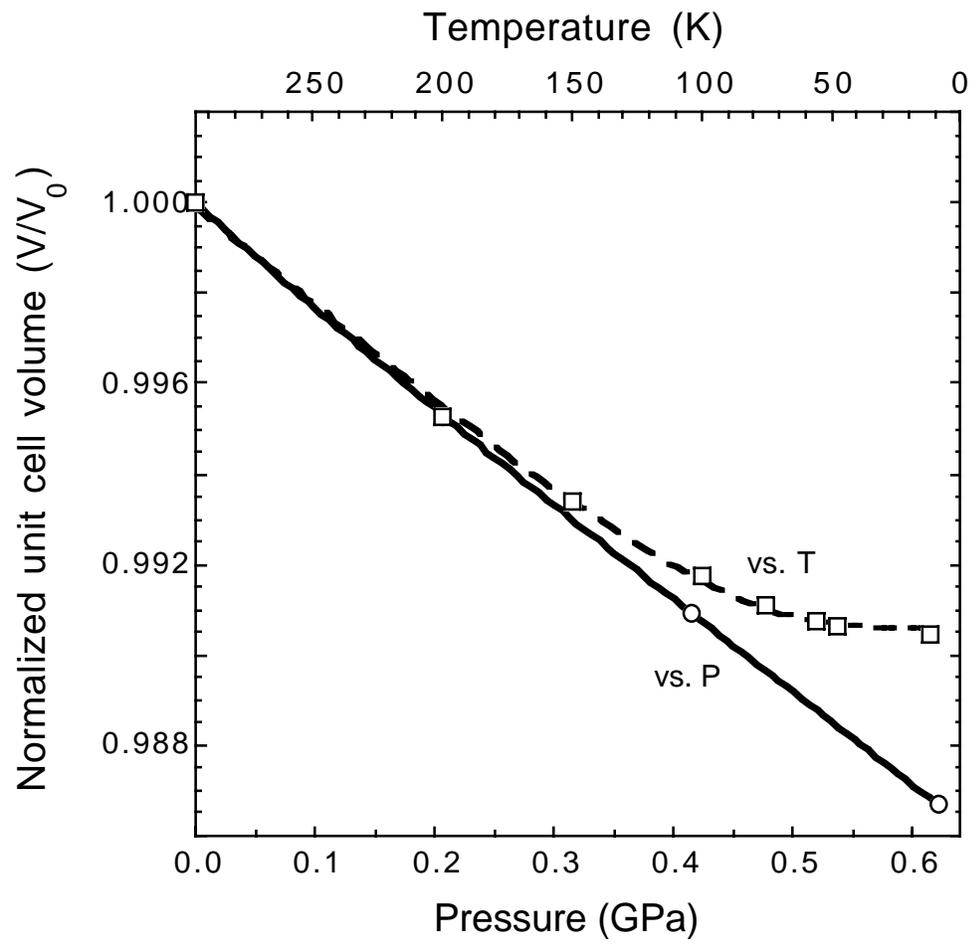

Fig. 1(b)

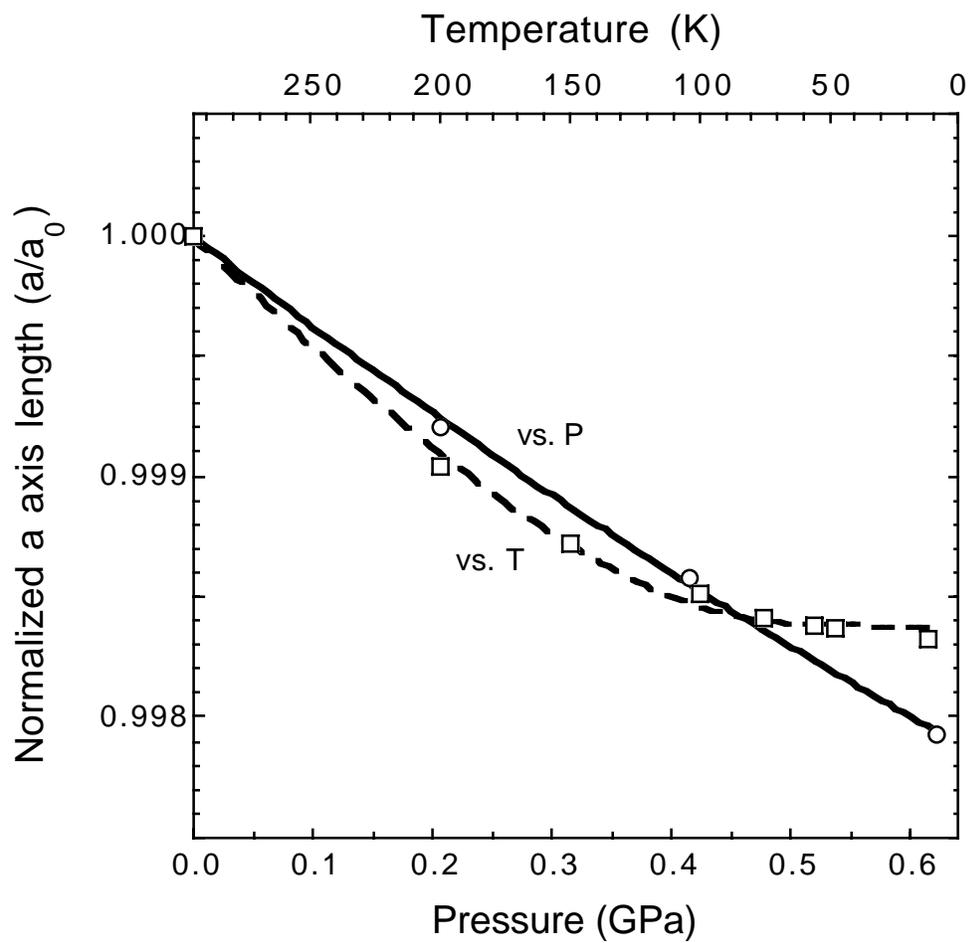

Fig. 1(c)

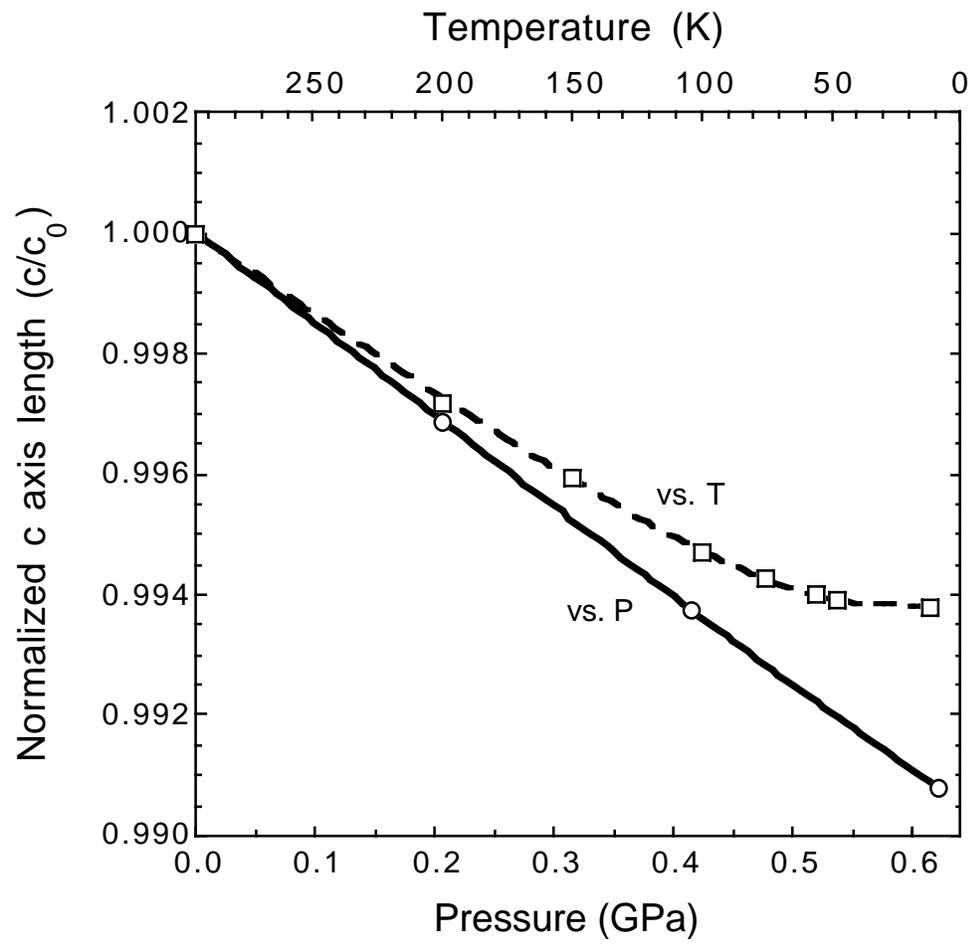

Fig. 2.

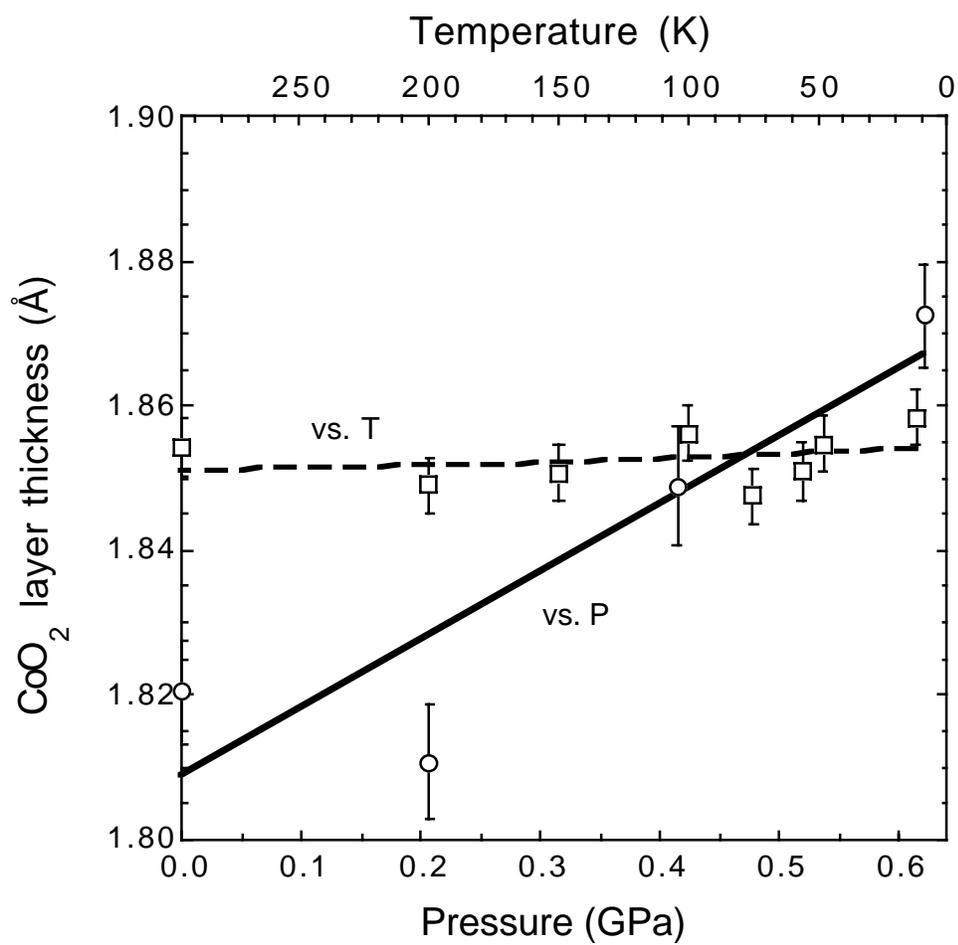

Fig. 3.

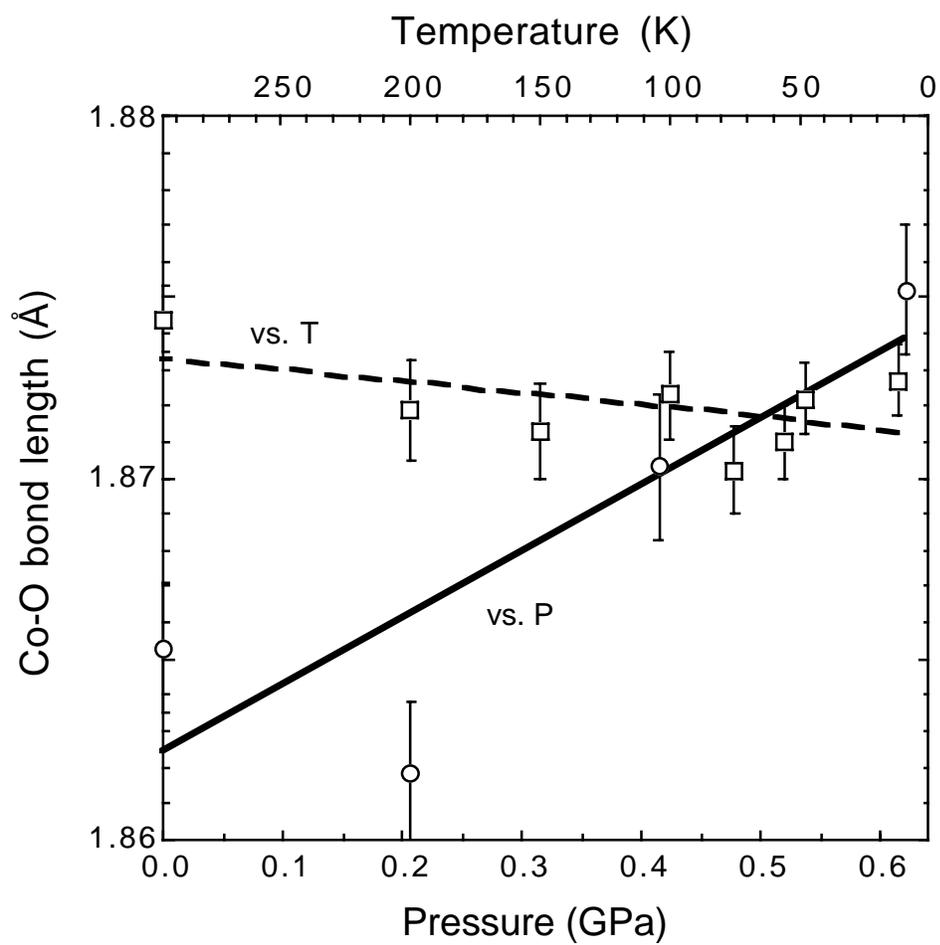

Fig. 4.

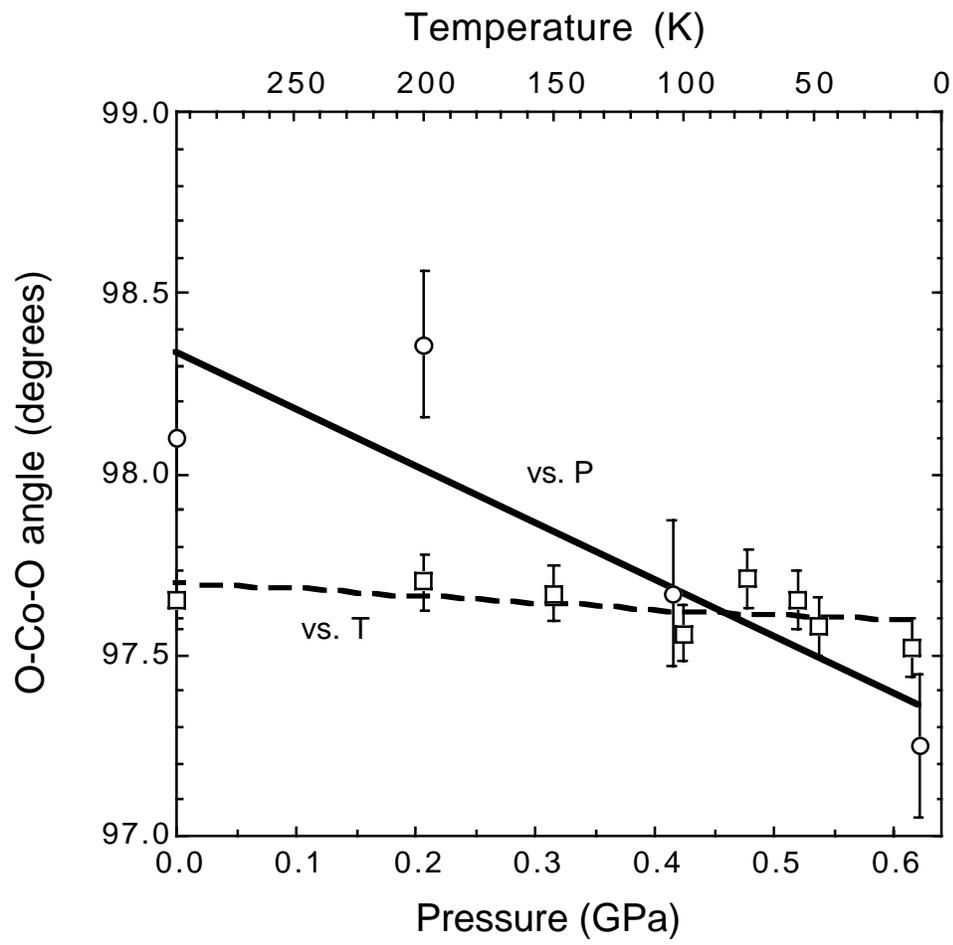

Fig. 5.

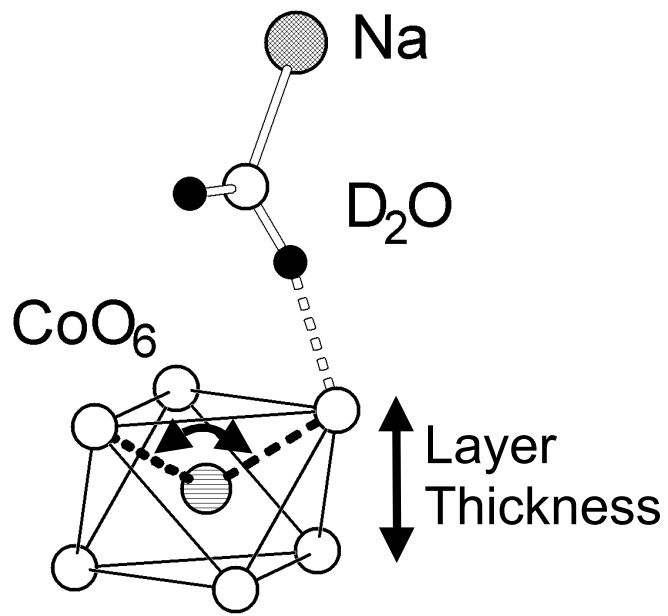